\documentclass[a4paper,12pt]{article}
\pdfoutput=1 %for pdflatex 

\usepackage{graphicx}
\usepackage{amsmath}
\usepackage{amssymb}
\usepackage{amsfonts}
\usepackage{amsthm}
\usepackage{xcolor}
\usepackage{cite}
\usepackage{tikz}

\newcommand{\del}{\partial}

 \makeatletter
 \@addtoreset{equation}{section}
 \makeatother

\newcommand{\Transpose}{\textsf{T}}
\newcommand{\dop}{\mathrm{d}}

\usepackage[bookmarks=true,bookmarksnumbered=true,bookmarkstype=toc]{hyperref} 
\hypersetup{
pdfauthor={Tetsuji KIMURA; Shin Sasaki; Kenta Shiozawa},
colorlinks={true},
linkcolor={black},
urlcolor={blue},
filecolor={black},
citecolor={blue}
}

\date{empty}
\begin{document}
\allowdisplaybreaks{
\begin{titlepage}
\null
\begin{flushright}
February, 2022
\end{flushright}
\vskip 2cm
\begin{center}
{\Large \bf 
Hyperk\"{a}hler, Bi-hypercomplex, 
\\
\vskip 0.5cm
Generalized Hyperk\"{a}hler Structures and T-duality 
}
\vskip 2cm
\normalsize
\renewcommand\thefootnote{\alph{footnote}}

{\large
Tetsuji Kimura${}^{*}$\footnote{t-kimura(at)osakac.ac.jp}, 
Shin Sasaki${}^{\dagger}$\footnote{shin-s(at)kitasato-u.ac.jp}
and 
Kenta Shiozawa${}^{\dagger}$\footnote{k.shiozawa(at)sci.kitasato-u.ac.jp}
}

\vskip 0.5cm

  {\it
  ${}^{*}$
  Center for Physics and Mathematics, Institute for Liberal Arts and
 Sciences, \\
  Osaka Electro-Communication University, Neyagawa, Osaka 572-8530,
 Japan \\
  \vspace{0.3cm}
  ${}^{\dagger}$
  Department of Physics,  Kitasato University \\
  Sagamihara 252-0373, Japan
  }
\vskip 1cm
\begin{abstract}
We exploit the doubled formalism to study
comprehensive relations among T-duality, complex and 
 bi-hermitian structures $(J_+, J_-)$ in two-dimensional $\mathcal{N} =(2,2)$
 sigma models with/without twisted chiral multiplets.
The bi-hermitian structures $(J_+,J_-)$ embedded in generalized
 K\"{a}hler structures $(\mathcal{J}_+,\mathcal{J}_-)$ are organized into
 the algebra of the tri-complex numbers.
We write down an analogue of the Buscher rule by which the T-duality
 transformation of the bi-hermitian and K\"{a}hler structures are apparent.
We also study the bi-hypercomplex and hyperk\"{a}hler cases in 
$\mathcal{N} = (4,4)$ theories.
They are expressed, as a T-duality covariant fashion, in the generalized hyperk\"{a}hler structures
and form the split-bi-quaternion algebras.
As a concrete example, we show the explicit T-duality relation between
the hyperk\"{a}hler structures of the KK-monopole (Taub-NUT space) and 
the bi-hypercomplex structures of the H-monopole (smeared NS5-brane).
Utilizing this result, we comment on a T-duality relation for the
 worldsheet instantons in these geometries. 
\end{abstract}
\end{center}

\end{titlepage}

\newpage
\setcounter{footnote}{0}
\renewcommand\thefootnote{\arabic{footnote}}
\pagenumbering{arabic}
\tableofcontents

%%%%%%%%%%%%%%%%%%%%%%%%%%%%%%%%%%%%%%%%%%%%%%%%%
\section{Introduction} \label{sect:introduction}
It is well-known that the complex structures in spacetime geometries appear in association with supersymmetries.
Target spaces of four-dimensional
$\mathcal{N} = 1$ and $\mathcal{N} = 2$ supersymmetric non-linear sigma models are described by
K\"{a}hler and hyperk\"{a}hler geometries, respectively \cite{Zumino:1979et,
Alvarez-Gaume:1980xat}.
The same is true for theories with equivalent numbers of supersymmetries.
For example, two-dimensional $\mathcal{N} = (2,2)$ and $\mathcal{N} =
(4,4)$ theories require K\"{a}hler and hyperk\"{a}hler geometries as target
spaces of chiral multiplets.
The geometries are characterized by complex structures $J$ or 
$J_a \, (a=1,2,3)$ compatible with the target space metric $g_{\mu \nu}$.
These facts are further generalized when the twisted chiral multiplets
come in the theories together.
For $\mathcal{N} = (2,2)$ theories with twisted chiral multiplets, the
target spaces admit bi-hermitian structures $(J_+,J_-)$ \cite{Gates:1984nk, Howe:1984fak}.
They are commutative complex structures and are compatible with the metric.
The geometries characterized by $(g_{\mu \nu}, J_+, J_-)$ are called
bi-hermitian manifolds.
For $\mathcal{N} = (4,4)$ cases, the target spaces are
bi-hypercomplex manifolds characterized by two commuting hypercomplex
structures $(J_{a,+}, J_{a,-}), \, (a=1,2,3)$ compatible with the metric.
These two-dimensional models may be identified with the worldsheet theories of fundamental strings on
certain background spacetimes.

This is not the end of the story.
Generalized geometry \cite{Hitchin:2003}, developed for understanding
the T-duality nature of target spaces $M$, plays an important role 
in relating these geometries.
In \cite{Gualtieri:2004}, it is shown that the bi-hermitian structures $(J_+,J_-)$ and the bi-hypercomplex structures 
$(J_{a,+}, J_{a,-})$
on the tangent bundle $TM$ are equivalent, via so-called the Gualtieri map, 
to the generalized K\"{a}hler structures $(\mathcal{J}_+, \mathcal{J}_-)$ 
and the generalised hyperk\"{a}hler structures 
$(\mathcal{J}_{a,+}, \mathcal{J}_{a,-})$
on the generalized tangent bundle $TM \oplus T^*M$, respectively.
These relations are also studied at the level of supersymmetric sigma models 
\cite{Lindstrom:2004eh, Lindstrom:2004iw, Bredthauer:2005zx, Lindstrom:2005zr}.
A physical origin of this correspondence comes from the equivalence of
the Lagrangian and the Hamiltonian formulations of sigma models \cite{Zabzine:2005qf, Bredthauer:2006hf}.
See \cite{Lindstrom:2012ci} and references therein for details.

One consequence of generalized geometry is that T-duality in string theory is realized in an apparent fashion.
This becomes obvious when its connection to double field theory (DFT) \cite{Hull:2009mi} is revealed.
DFT is developed from the doubled formalism \cite{Siegel:1993th, Siegel:1993xq} where T-duality is realized manifestly.
The dynamical fields in DFT are the generalized metric $\mathcal{H}_{MN}$ and the generalized dilaton $d$.
They are subject to physical conditions known as the weak and the strong constraints. With these constraints
, the spacetime metric $g_{\mu \nu}$, the NSNS $B$-field
$B_{\mu \nu}$ and the dilaton $\phi$ are nicely packaged in $\mathcal{H}_{MN}$.
They are defined on a $2D$-dimensional 
para-K\"{a}hler or para-hermitian manifold $\mathcal{M}$ \cite{Vaisman:2012ke, Vaisman:2012px, Marotta:2018myj}, 
that are sometimes called doubled space, 
where the Kaluza-Klein and the winding coordinates $x^M = (x^{\mu},
\tilde{x}_{\mu})$ are naturally introduced.
The famous Buscher rule of T-duality is realized as an $O(D,D)$ symmetry on 
the doubled tangent space $T \mathcal{M}$.
Due to the para-hermitian structure, $T\mathcal{M}$ is decomposed into two transversal parts $L \oplus \tilde{L}$.
The physical $D$-dimensional spacetime in $\mathcal{M}$ is defined by
the foliated space $M$ for $L = TM$.
The generalized and the doubled geometries are different since the
former is build on the $D$-dimensional base space $M$ while the latter
assumes the $2D$-dimensional doubled base space $\mathcal{M}$.
Although they are different, they are conceptually identified in the
following sense.
The weak and the strong constraints are trivially solved by
$\mathcal{H}_{MN}$ and $d$ that depend on a half of the doubled
coordinate $x^M$. This parameterizes the foliated space $M$.
In other words, the constraints in DFT restrict the base space $\mathcal{M}$ to
a $D$-dimensional subspace $M$ while keeping $T\mathcal{M}$ intact.
With this perception, $T \mathcal{M}$ is identified with 
the generalized tangent bundle $TM \oplus T^*M$ through 
so-called the natural isomorphism \cite{Vaisman:2012ke, Freidel:2017yuv, Freidel:2018tkj}.
In this sense, generalized geometry is implemented within the doubled geometry.
Indeed, the $O(D,D)$ is the structure group of the generalized tangent
bundle $TM \oplus T^*M$.

With these facts at hand, it is now legitimate to discuss the 
relation between 
the K\"{a}hler (hyperk\"{a}hler) structures and the bi-hermitian (bi-hypercomplex) 
from the viewpoint of T-duality.
Since the chiral and the twisted chiral multiplets are
interchanged by T-duality \cite{Lindstrom:1983rt, Gates:1983nr, Rocek:1991ps}, 
there should be an explicit transformation rule 
from the K\"{a}hler (hyperk\"{a}hler) structures to the bi-hermitian (bi-hypercomplex), and vice versa, 
as target spaces of two-dimensional sigma models.
In this paper, we establish these relations by working in the doubled
formalism.
We also note that the bi-hermitian (bi-hypercomplex) and the
generalized (hyper)k\"{a}hler structures involve interesting mathematical
properties and they are relevant to T-duality of worldsheet instanton effects.

The organization of this paper is as follows.
In the next section, we provide a brief overview on the two-dimensional
$\mathcal{N} = (2,2)$ and $\mathcal{N} = (4,4)$ sigma models and
introduce the bi-hermitian (bi-hypercomplex) structures in their target spaces.
We then discuss the embedding of these structures into the generalized
complex and the hyperk\"{a}hler structures on $TM \oplus T^*M$.
We analyze hypercomplex algebras that 
the structures obey. 
In Section \ref{section:T-duality}, we consider the
T-duality transformations of the generalized complex structures.
Using this, we extract the T-duality transformation of the bi-hermitian
structures and write down the explicit ``Buscher rule'' of T-duality for
them.
In order to get a better understanding of the relation between the 
hyperk\"{a}hler and bi-hypercomplex structures, Section \ref{section:HM_KKM} is devoted to an explicit
example of T-duality for these structures.
We focus on the T-duality between the 
KK- and the H-monopoles and show how
the hyperk\"{a}hler and the bi-hypercomplex 
structures are related.
As a byproduct, we comment on the T-duality between the
worldsheet instanton equations in these geometries in Section \ref{section:instantons}.
Section \ref{section:conclusion} is devoted to conclusion and discussions.

\section{$\mathcal{N} = (2,2)$, $\mathcal{N} = (4,4)$ sigma models and generalized (hyper)K\"{a}hler structures} \label{section:NLSM}
In this section, we give a brief overview on 
two-dimensional $\mathcal{N} = (2,2)$ and $\mathcal{N} = (4,4)$ 
sigma models with chiral and twisted chiral multiplets and their target space geometries.
The structures of the target spaces are well described by generalized
geometry or doubled geometry. We introduce these notions in the following.
For details, see \cite{Gates:1984nk} and references therein. 

\subsection{Supersymmetric sigma models and their target space geometries}

The most general action containing two-dimensional $\mathcal{N} = (2,2)$
chiral superfields 
$\hat{\Phi}^u \, (u=1,\ldots,n)$ 
and twisted chiral superfields 
$\hat{\chi}^p \, (p=1,\ldots, m)$ 
is given by 
\begin{align}
S = \int {\dop}^2 x {\dop}^2 \theta {\dop}^2 \bar{\theta} \, K (\hat{\Phi}^u, \bar{\hat{\Phi}}^{\bar{v}},
 \hat{\chi}^p, \bar{\hat{\chi}}^{\bar{q}}).
\label{eq:2d_NLSM_action}
\end{align}
Here $K$ is a real function. The action \eqref{eq:2d_NLSM_action} is
invariant under the following generalized K\"{a}hler transformation;
\begin{align}
\delta K = \Lambda_1 (\hat{\Phi}, \hat{\chi}) + \Lambda_2 (\hat{\Phi}, \bar{\hat{\chi}}) + \bar{\Lambda}_1 (\bar{\hat{\Phi}}, \bar{\hat{\chi}}) + \bar{\Lambda}_2 (\bar{\hat{\Phi}},\hat{\chi}).
\label{eq:generalized_Kahler_transformation}
\end{align}
The bosonic part of the action \eqref{eq:2d_NLSM_action} is found to be
\begin{align}
S =& \ - \frac{1}{2} \int \! {\dop}^2 x \, 
\Big[
K_{u \bar{v}} \del_a \varphi^u \del^a \bar{\varphi}^{\bar{v}}
-
K_{p \bar{q}} \del_a \chi^p \del^a \bar{\chi}^{\bar{q}}
+
\varepsilon^{ab} 
\left(
K_{u \bar{p}} \del_a \varphi^u \del_b \bar{\chi}^{\bar{p}}
+
K_{p \bar{u}} \del_a \bar{\varphi}^{\bar{u}} \del_b \chi^p
\right)
\Big],
\label{eq:component_action}
\end{align}
where $\varphi^u$ and $\chi^p$ are the lowest components in the
superfields 
$\hat{\Phi}^u$ and $\hat{\chi}^p$, respectively.
We have also defined $K_{u \bar{v}} = \frac{\del^2 K}{\del \varphi^u
\del \bar{\varphi}^{\bar{v}}}$ and so on.
In order that the theory is ghost free, we demand that $K_{p\bar{q}}$ is
negative-definite. 
Since the kinetic and the Wess-Zumino terms of the scalar fields define the
metric and the $B$-field in the target space, 
the presence of the twisted chiral multiplets is necessary to
introduce the $B$-field and hence the torsion is given by $T = \dop B$.

We next examine supersymmetric transformations.
Following \cite{Gates:1984nk}, we first write down the action in
terms of $\mathcal{N} = (1,1)$ real superfields $\Phi^{\mu} \,
(\mu=1,\ldots, D)$;
\begin{align}
S = - \frac{1}{4} \int \! {\dop}^2 x {\dop}^2 \theta \,
\Big[
g_{\mu \nu} (\Phi) D^{\alpha} \Phi^{\mu} D_{\alpha} \Phi^{\nu} + B_{\mu \nu} (\Phi) D^{\alpha} \Phi^{\mu} (\gamma_5 D)_{\alpha} \Phi^{\nu}
\Big].
\label{eq:2d_N1}
\end{align}
Here $D_{\alpha} \, (\alpha = \pm)$ is 
$\mathcal{N} = (1,1)$ supercovariant derivative, 
$g_{\mu \nu}$, $B_{\mu \nu} = - B_{\nu \mu}$ are the metric and the $B$-field in 
the target space and $\gamma_5 = \sigma^3$. 
The action is invariant under the gauge
transformation $\delta B = \dop \xi$. In the following, the torsion in the
target space is always given by $T = \dop B$.
The action \eqref{eq:2d_N1} is manifestly invariant under the
$\mathcal{N} = (1,1)$ supersymmetry transformation.
We further require 
that the action is invariant under the following
additional supersymmetry transformation;
\begin{align}
\delta_{\eta} \Phi^{\mu} = 
- i 
(J_+)^\mu{}_\nu 
\left(
\eta_{+} D_{-} \Phi^{\nu}
\right) 
+ i
(J_-)^\mu{}_\nu
\left(
\eta_{-} D_+ \Phi^{\nu}
\right),
\label{eq:add_SUSY}
\end{align}
where $\eta_{\pm}$ are supersymmetry parameters and $(J_\pm)^\mu{}_\nu (\Phi)$
are some matrices.
By demanding that the commutator of the above transformations closes to
translations on-shell, we find a condition on $\Phi^{\mu}$;
\begin{align}
D_+ D_- \Phi^{\mu} + \Gamma^{(\pm)\mu} {}_{\nu \rho} D_+ \Phi^{\nu} D_- \Phi^{\rho} = 0,
\end{align}
where 
$\Gamma^{(\pm) \mu} {}_{\nu \rho} = \Gamma^{(0)\mu} {}_{\nu \rho} \mp T^{\mu}{}_{\nu \rho}$
is the affine connection with the torsion $T$ and 
\begin{align}
\Gamma^{(0)\mu} {}_{\nu \rho} = \frac{1}{2} g^{\mu \sigma} 
\left(
\del_{\rho} g_{\nu \sigma} + \del_{\nu} g_{\rho \sigma} - \del_{\sigma} g_{\nu \rho}
\right)
\end{align}
is the Levi-Civita connection. 
By imposing the supersymmetry algebra, we find conditions on $J_{\pm}$;
\begin{align}
(J_{\pm})^{\mu}{}_{\nu} (J_{\pm})^{\nu}{}_{\rho}
&= - \delta^{\mu}{}_{\rho},
\notag \\
(N_{\pm})_{\mu \nu}{}^{\rho}
&= 
(J_{\pm})^\sigma{}_{\mu} \del_{[\sigma|} (J_{\pm})^\rho{}_{|\nu]}
-
(J_{\pm})^\sigma{}_{\nu} \del_{[\sigma|} (J_{\pm})^\rho{}_{|\mu]}
= 0.
\end{align}
Here $N_{\pm}$ are the Nijenhuis tensors associated with $J_{\pm}$ whose
vanishing condition implies the integrability of $J_{\pm}$.
Then, $J_{\pm}$ become the complex structures on the target space.

From the invariance of the action by the transformation
\eqref{eq:add_SUSY}, one finds the additional relations;
\begin{align}
g_{\mu\rho} (J_{\pm})^{\rho}{}_{\nu} 
&= - g_{\nu\rho} (J_{\pm})^\rho{}_\nu,
\notag \\
\nabla^{(\pm)}_{\mu} 
(J_{\pm})^{\nu}{}_{\rho}
&=
\del_{\mu} (J_{\pm})^{\nu}{}_{\rho} 
+ (J_\pm)^\sigma{}_\rho \Gamma^{(\pm)\nu}{}_{\mu\sigma}
- (J_\pm)^\nu{}_\sigma \Gamma^{(\pm)\sigma}{}_{\mu\rho}
= 0.
\label{eq:covariantly_constant}
\end{align}
They declare that the metric is hermitian with respect to $J_{\pm}$. 
The complex structures are covariantly constant for the affine connections $\Gamma^{(\pm)}$.
Note that the metric is covariantly constant with respect to both
$\Gamma^{(\pm)}$.
From the integrability of $J_{\pm}$ and the fact that they are covariantly
constant, we have a non-trivial constraint on $J_{\pm}$;
\begin{align}
(J_+)_{\mu \nu \rho} (J_+)^\mu{}_{\sigma} (J_+)^\nu{}_{\tau} (J_+)^\rho{}_{\kappa}
=
- (J_-)_{\mu \nu \rho} (J_-)^\mu{}_{\sigma} (J_-)^\nu{}_{\tau} (J_-)^\rho{}_{\kappa}.
\label{eq:I_constraint}
\end{align}
Here $(J_{\pm})_{\mu \nu \rho} = \frac{1}{2} \del_{[\rho} (J_{\pm})_{\mu \nu]}$ and
$(J_{\pm})_{\mu \nu} = - g_{\mu\rho} (J_+)^\rho{}_{\nu}$.

Now we demand that the two complex structures commute with each other;
\begin{align}
[J_+, J_-] = 0.
\end{align}
Then, we can define the following almost product structure;
\begin{align}
\Pi^{\mu}{}_{\nu} = 
%- 
(J_+)^{\mu}{}_{\rho} (J_-)^{\rho}{}_{\nu}.
\end{align}
By definition, $\Pi$ commutes with $J_{\pm}$ and satisfies the relation
\begin{align}
\Pi^{\mu}{}_{\rho} \Pi^{\rho}{}_{\nu} = \delta^{\mu}{}_{\nu}.
\end{align}
Therefore $\Pi$ is regarded as the almost real structure.
One can show that when $J_{\pm}$ are integrable, then $\Pi$ is also.
By the integrability of $\Pi$ and the hermiticity of 
$g_{\mu \nu}$ with respect to $J_{\pm}$, we find the relation
\begin{align}
g_{\mu\rho} \Pi^\rho{}_\nu 
= g_{\nu\rho} \Pi^\rho{}_{\mu}.
\end{align}
With these structures, we can choose 
$g_{\mu \nu}$ as a block diagonal form. 
By using
\eqref{eq:I_constraint}, we obtain
\begin{align}
g_{p \bar{q}} = \frac{\del^2 K}{\del \Phi^p \del \bar{\Phi}^{\bar{q}}},
\quad
g_{u \bar{v}} = - \frac{\del^2 K}{\del \Phi^u \del \bar{\Phi}^{\bar{u}}}.
\end{align}
Here $K$ is the real function appeared in \eqref{eq:2d_NLSM_action}
and we have decomposed the indices $\mu,\nu = 1, \ldots, D = 2 (m+n)$ into 
$p,q = 1, \ldots, m$ and $u,v = 1, \ldots, n$.
The torsion is also determined by the function $K$ and the action
\eqref{eq:component_action} is recovered.
This $K$ is determined uniquely up to the generalized K\"{a}hler
transformation \eqref{eq:generalized_Kahler_transformation}.
From these structures we can read off the supersymmetry transformations;
\begin{align}
\delta_{\eta} \Phi^u
&= J^u{}_v (\eta^{\alpha} D_{\alpha} \Phi^v),
\notag \\
\delta_{\eta} \Phi^p 
&= \tilde{J}^p{}_q (\eta^{\alpha} (\gamma_5 D)_{\alpha} \Phi^q),
\end{align}
where $J_{\pm} = \tilde{J} \pm J$.
The superfields $\Phi^u, \, (u=1, \ldots, n)$ and 
$\Phi^p, \, (p=1, \ldots, m)$ are the $\mathcal{N} = (1,1)$ components of the 
$\mathcal{N} = (2,2)$ chiral superfields $\hat{\Phi}$
and the twisted chiral superfields $\hat{\chi}$.
Note that $K$ is not a K\"{a}hler potential and the target space 
is not a K\"{a}hler manifold anymore.
Since the target space has the two hermitian structures associated with
the two complex structures $J_{\pm}$, 
this is called a bi-hermitian manifold.

In summary, the bi-hermitian structures are introduced as a way to realize
the $\mathcal{N} = (2,2)$ supersymmetry for theories with chiral and twisted chiral multiplets.
In this case, the target manifold admits the metric 
$g_{\mu \nu}$, the $B$-field $B_{\mu \nu}$ and the torsion associated
with the $B$-field $T = {\dop}B$. 
The bi-hermitian manifold is defined by two commuting complex
structures $J_{\pm}$, and they are compatible with the metric 
$g(J_{\pm} \cdot, J_{\pm} \cdot) = g(\cdot, \cdot)$. 
Obviously, when the two complex structures coincide $J_+ = J_-$, the
geometry becomes a K\"{a}hler manifold.
This is realized only when the $B$-field is pure gauge and the torsion vanishes.

The same is true for $\mathcal{N} = (4,4)$ chiral and twisted chiral multiplets. 
The $\mathcal{N} = (4,4)$ supersymmetry requires the complex
structures $J_{a,\pm} \, (a=1,2,3)$ satisfying the relations;
\begin{align}
&
(J_{a,\pm})^{\mu}{}_{\nu} (J_{b,\pm})^{\nu}{}_{\rho} 
+ (J_{b,\pm})^{\mu}{}_{\nu} (J_{a,\pm})^{\nu}{}_{\rho} 
= - 2 \delta_{ab} \, \delta^{\mu}{}_{\rho},
\quad 
(a,b = 1,2,3),
\notag \\
&
[J_{a,+}, J_{b,-}] = 0.
\end{align}
Each $J_{a,\pm}$ satisfies the quaternionic structures and the
dimension of the target space is $4k$. 
The metric is hermitian with respect to $J_{a,\pm}$.
Since the target space admits two commuting hyperk\"{a}hler structures, 
the geometry is called a bi-hypercomplex manifold\footnote{
This is also known as a hyperk\"{a}hler with torsion geometry in vast literature.
}.
In this case, there are $3 \times 3 = 9$ almost product structures
\begin{align}
\Pi_{ab} = J_{a,+} J_{b,-}.
\end{align}
Similar to $\mathcal{N} = (2,2)$ cases, an $\mathcal{N} = (4,4)$ theory
is governed by a real function $K$.
When each complex structure coincides 
$J_{a,+} = J_{a,-} \, (a=1,2,3)$ with each other,
the geometry becomes a hyperk\"{a}hler manifold.

Before we move to the next section, a comment is in order.
We have discussed the bi-hermitian and the bi-hypercomplex structures
where the complex structures $J_+,J_-$ commute with each other.
However, this is not generically satisfied.
It is known that when there are semi-chiral multiplets in sigma models,
$J_+, J_-$ cease to be commuting and the bi-hermitian (bi-hypercomplex)
geometry admits more general structures 
\cite{Lindstrom:2004hi, Lindstrom:2005zr, Lindstrom:2007qf,
Goteman:2009ye, Goteman:2012qk, Lindstrom:2014bra}.
In the following, we never assume the semi-chiral multiplets and
consider commuting complex structures.

\subsection{Generalized geometries}
The bi-hermitian and the bi-hypercomplex structures of spacetime $M$
are well described in the context of generalized geometry or doubled geometry.

The generalized tangent bundle $\mathbb{T}M = TM \oplus T^*M$, or equivalently, the
doubled tangent bundle $T\mathcal{M}$ is a natural arena to discuss
T-duality relations of various quantities.
For example, the metric $g_{\mu \nu}$, the $B$-field $B_{\mu \nu}$ and the dilation $\phi$ are organized into the
generalized metric $\mathcal{H}_{MN}$ and the generalized dilation $d$;
\begin{align}
\mathcal{H}_{MN} = 
\left(
\begin{array}{cc}
g_{\mu \nu} - B_{\mu \rho} g^{\rho \sigma} B_{\sigma \nu} & B_{\mu \rho}
 g^{\rho \nu}  \\
- g^{\mu \rho} B_{\rho \nu} & g^{\mu \nu}
\end{array}
\right), \qquad 
e^{-2d} = \sqrt{-g} e^{-2\phi},
\end{align}
where $M,N = 1, \ldots, 2D$.
The T-duality transformations of $g_{\mu \nu}, B_{\mu \nu}$ and $\phi$
are read off from the $O(D,D)$ transformation of $\mathcal{H}_{MN}$ and
$d$;
\begin{align}
\mathcal{H}'_{MN} = \mathcal{O}^{\Transpose}_M {}^P \mathcal{H}_{PQ} \mathcal{O}^Q
 {}_N, \qquad
e^{-2d'} = e^{-2d}.
\end{align}
Here the generalized dilation $d$ is invariant under the $O(D,D)$ rotation.
Explicitly, the famous Buscher rule of the T-duality transformation \cite{Buscher:1987sk}
\begin{align}
g'_{ij} &=  g_{ij} - {g_{iy} g_{jy} - B_{iy} B_{jy} \over g_{yy}}, &\qquad
g'_{iy} &=  {B_{iy} \over g_{yy}}, & \qquad
g'_{yy} &=  {1 \over g_{yy}}, 
\notag \\
B'_{ij} &=  B_{ij} - {B_{iy} g_{jy} - g_{iy} B_{jy} \over g_{yy}}, &\qquad
B'_{iy} &=  {g_{iy} \over g_{yy}}. & \phi' &= \phi - \frac{1}{2} \log
 g_{yy}, & (i,j \not= y),
\label{eq:Buscher_rule}
\end{align}
are obtained by the factorized $O(D,D)$ T-duality transformation;
\begin{align}
\mathcal{O} = h_y = 
\left(
\begin{array}{cc}
1 - t_y & t_y  \\
t_y & 1 - t_y
\end{array}
\right),
\qquad 
(t_y)^{\mu} {}_{\nu} = \delta^{\mu}_y \delta^y_{\nu},
\qquad
h_y^{\Transpose} = h_y.
\label{eq:factorized_T-dual}
\end{align}
Here $y$ is the isometry direction where the T-duality transformation is
performed.

Just as $g_{\mu \nu}, B_{\mu \nu}, \phi$ in the spacetime $M$ are expressed 
in a T-duality covariant way, the complex structure $J$ 
and the K\"{a}hler form $\omega = - g J$ are also given in the doubled formalism.
A generalized almost complex structure $\mathcal{J}$ is defined by an
endomorphism $\mathcal{J} : \mathbb{T}M \to \mathbb{T}M$ that 
preserves the inner product on $\mathbb{T}M$ and
squares to the minus identity $\mathcal{J}^2 = - \mathbf{1}_{2D}$. 
For simplicity, we consider $B_{\mu \nu} = 0$ for the time being.
Given the structures $J$ and $\omega$ in
$M$, one shows that 
\begin{align}
\mathcal{I}_J = 
\left(
\begin{array}{cc}
J & 0  \\
0 & - J^*
\end{array}
\right), 
\qquad 
\mathcal{I}_{\omega} = 
\left(
\begin{array}{cc}
0 & - \omega^{-1} \\
\omega & 0
\end{array}
\right)
\end{align}
are generalized almost complex structures on $\mathbb{T}M$.
Here $J^* : T^*M \to T^*M$ is the adjoint of $J$.
%When $J$ and $\omega$ are integrable, 
When $J$ is integrable and $\omega$ is closed,
namely, the Nijenhuis tensor for $J$ vanishes and $\dop \omega = 0$, 
then $\mathcal{I}_J$ and $\mathcal{I}_{\omega}$ are 
Courant integrable, respectively \cite{Gualtieri:2004}. 
Here the Courant integrability is defined by the involutivity of the 
$+i$-eigen bundle by $\mathcal{I}_J, \mathcal{I}_{\omega}$ for the
Courant bracket.
In this case, the generalized almost complex
structures become generalized complex structures.

A generalized K\"{a}hler structure is defined by a pair of 
commutative 
generalized complex structures $(\mathcal{J}_1, \mathcal{J}_2)$
whose product $\mathcal{G} = \mathcal{J}_1 \mathcal{J}_2$ defines a positive-definite metric on $\mathbb{T}M$.
One easily finds that 
$\mathcal{I}_J$ commutes with $\mathcal{I}_{\omega}$ and 
$\mathcal{G} = \mathcal{I}_J \mathcal{I}_{\omega}$ is positive-definite
on $\mathbb{T}M$. Then $(\mathcal{I}_J, \mathcal{I}_{\omega})$ is 
the generalized K\"{a}hler structure.
Since $\mathcal{I}_J$ 
commutes with $\mathcal{I}_{\omega}$, 
it is obvious that $\mathcal{G}$ is the real structure on
$\mathbb{T}M$. 
The structures $(\mathcal{I}_J, \mathcal{I}_{\omega}, \mathcal{G})$
together with the identity $\mathbf{1}_{2D}$ on $\mathbb{T}M$ form 
the algebra of the bi-complex numbers $\mathbb{C}_2$; 
\begin{align}
\mathcal{I}_J^2
 = \mathcal{I}_{\omega}^2 = -\mathbf{1}_{2D}, 
\quad 
\mathcal{G}^2 = \mathbf{1}_{2D},
 \quad 
\mathcal{I}_J \mathcal{I}_{\omega} \mathcal{G} = \mathbf{1}_{2D}.
\label{eq:bi-complex}
\end{align}

We can also show that the bi-hermitian structures define generalized complex structures;
\begin{align}
\mathcal{J}_{\pm} = \frac{1}{2} 
\left(
\begin{array}{cc}
J_+ \pm J_- & - (\omega_+^{-1} \mp \omega^{-1}_-) \\
\omega_+ \mp \omega_- & - (J^*_+ \pm J^*_-)
\end{array}
\right),
\label{eq:Gualtieri_map}
\end{align}
where $\omega_{\pm} = - g J_{\pm}$ are the fundamental two-forms associated with the bi-hermitian structures $J_{\pm}$.
Since they commute with each other $[\mathcal{J}_+, \mathcal{J}_-] = 0$, 
they define the generalized K\"{a}hler structure.
In this sense, the set
$(J_+, J_-)$ together with the metric $g_{\mu \nu}$ 
is equivalent to that in $(\mathcal{J}_+, \mathcal{J}_-)$.
The map \eqref{eq:Gualtieri_map} is sometimes called the Gualtieri map.
The physical origin of this correspondence is studied
\cite{Lindstrom:2004eh, Lindstrom:2004iw, Bredthauer:2005zx,
Lindstrom:2005zr, Zabzine:2005qf, Bredthauer:2006hf}.
We find that each $(\mathcal{I}_{J_+}, \mathcal{I}_{\omega_+}, \mathcal{G})$ 
and $(\mathcal{I}_{J_-}, \mathcal{I}_{\omega_-}, \mathcal{G})$ forms the
algebra of the bi-complex numbers.
Here $\mathcal{G} = \mathcal{I}_{J_+} \mathcal{I}_{\omega_+} = \mathcal{I}_{J_-} \mathcal{I}_{\omega_-}$ is 
the common real structure.
Furthermore, we find that there are additional real structures on $\mathbb{T}M$;
\begin{align}
\mathcal{G}' = \mathcal{I}_{J_+} \mathcal{I}_{J_-}
= \mathcal{I}_{\omega_+} \mathcal{I}_{\omega_-},
\qquad
\mathcal{G}'' = \mathcal{I}_{\omega_+} \mathcal{I}_{J_-}
= \mathcal{I}_{\omega_-} \mathcal{I}_{J_+}.
\end{align}
Altogether, once again by incorporating the identity 
$\mathbf{1}_{2D}$ on $\mathbb{T}M$, we obtain four real and four complex structures on
$\mathbb{T}M$ from the
given bi-hermitian structures on $M$.
They obey an eight-dimensional algebra of so-called the tri-complex
numbers $\mathbb{C}_3$.

The discussion is parallel in the bi-hypercomplex cases.
One finds that the bi-hypercomplex structures $(J_{a,+}, J_{a,-})$ on
$M$ defines generalized complex structures;
\begin{align}
\mathcal{J}_{a, \pm} = \frac{1}{2} 
\begin{pmatrix}
J_{a,+} \pm J_{a,-} & - (\omega_{a,+}^{-1} \mp \omega^{-1}_{a,-}) \\
\omega_{a,+} \mp \omega_{a,-} & - (J^*_{a,+} \pm J^*_{a,-})
\end{pmatrix}, 
\qquad (a=1,2,3),
\label{eq:generalized_hyperkahler_structures}
\end{align}
where $J_{a,\pm}$ are two commuting hyperk\"{a}hler structures on $M$ and
$\omega_{a,\pm} = - g J_{a,\pm}$ are the associated fundamental two-forms.
One finds that $\mathcal{J}_{a,\pm}$ in
\eqref{eq:generalized_hyperkahler_structures} satisfy the following relations;
\begin{align}
&
\mathcal{J}_{a,+} \mathcal{J}_{b,+} = - \delta_{ab} \mathbf{1}_{2D} 
+ \epsilon_{abc} \mathcal{J}_{c,+},
&&
\mathcal{J}_{a,-} \mathcal{J}_{b,-} = - \delta_{ab} \mathbf{1}_{2D} 
+ \epsilon_{abc} \mathcal{J}_{c,+},
\notag \\
&
\mathcal{J}_{a,+} \mathcal{J}_{b,-} =  \delta_{ab} \mathcal{G} 
+ \epsilon_{abc} \mathcal{J}_{c,-},
&&
\mathcal{J}_{a,-} \mathcal{J}_{b,+} =  \delta_{ab} \mathcal{G} 
+ \epsilon_{abc} \mathcal{J}_{c,-}.
\label{eq:generalized_hyperkahler}
\end{align}
The generalized complex structures satisfying the relations
\eqref{eq:generalized_hyperkahler} are known as 
the generalized hyperk\"{a}hler structures \cite{Bredthauer:2006sz}.
We find that the algebra \eqref{eq:generalized_hyperkahler} is that of
the split-bi-quaternions.

As we have discussed, the bi-hermitian and K\"{a}hler geometries are
described by $\mathcal{N} = (2,2)$ models with and without twisted
chiral multiplets. Since the chiral and the twisted chiral multiplets
are interchanged by T-duality, 
we next examine the explicit transformation rule for these geometries.

\section[T-duality transformation between two different bi-hermitian structures]{T-duality transformation 
between two different \\ bi-hermitian structures} \label{section:T-duality}
We start by the generalized complex structures 
associated with the bi-hermitian structures;
\begin{align}
\mathcal{J}_{\pm}
&= {1 \over 2} 
e^B
\begin{pmatrix}
J_+ \pm J_- & - (\omega_+^{-1} \mp \omega_-^{-1}) \\
\omega_+ \mp \omega_- & - (J_+^* \pm J_-^*) 
\end{pmatrix}
e^{-B},
\qquad
e^B =
\begin{pmatrix}
1 & 0 \\
B & 1
\end{pmatrix}
\in O(D,D),
\label{eq:general_parametrization_of_generalized_complex_structures}
\end{align}
where we have performed the $B$-transformation $e^B$ to 
include the $B$-field.
Note that in this case, the integrability of the generalized complex structures are
defined by the $H$-twisted Courant bracket \cite{Gualtieri:2004}.
One finds that the expression
\eqref{eq:general_parametrization_of_generalized_complex_structures} is
decomposed as
\begin{align}
\mathcal{J}_+
&= {1 \over 2} e^B \Big(
	\mathcal{I}_{J_+}
	+ \mathcal{I}_{J_-}
	+ \mathcal{I}_{\omega_+}
	- \mathcal{I}_{\omega_-} \Big) e^{-B}, 
\\
\mathcal{J}_-
&= {1 \over 2} e^B \Big(
	\mathcal{I}_{J_+}
	- \mathcal{I}_{J_-}
	+ \mathcal{I}_{\omega_+}
	+ \mathcal{I}_{\omega_-} \Big) e^{-B}.
\end{align}
Here
$\mathcal{I}_{J_{\pm}}$ and $\mathcal{I}_{\omega_{\pm}}$ are the
generalized complex structures of the form;
\begin{align}
\mathcal{I}_{J_\pm}
&= 
\begin{pmatrix}
J_\pm & 0 \\
0 & - J_\pm^*
\end{pmatrix}, \qquad
\mathcal{I}_{\omega_\pm}
=
\begin{pmatrix}
0 & - \omega_\pm^{-1} \\
\omega_\pm & 0
\end{pmatrix}.
\end{align}

We now perform the T-duality transformation along the $y$-direction.
The generalized complex structures $\mathcal{J}_+$ and $\mathcal{J}_-$ 
are then transformed as 
\begin{align}
\mathcal{J}'_+
&= h_y \mathcal{J}_+ h_y
= {1 \over 2} h_y e^B \Big(
	\mathcal{I}_{J_+}
	+ \mathcal{I}_{J_-}
	+ \mathcal{I}_{\omega_+}
	- \mathcal{I}_{\omega_-} \Big) e^{-B} h_y, 
\label{eq:T-dual_genKaehler_plus_1} \\
\mathcal{J}'_-
&= h_y \mathcal{J}_- h_y
= {1 \over 2} h_y e^B \Big(
	\mathcal{I}_{J_+}
	- \mathcal{I}_{J_-}
	+ \mathcal{I}_{\omega_+}
	+ \mathcal{I}_{\omega_-} \Big) e^{-B} h_y.
\label{eq:T-dual_genKaehler_minus_1}
\end{align}
Here, $h_y \in O(D,D)$ is the factorized T-duality transformation given
in \eqref{eq:factorized_T-dual}.
On the other hand, 
the generalized complex structures 
$\mathcal{J}'_\pm$
after the T-duality transformation 
are parameterized in such a way that
\begin{align}
\mathcal{J}'_+
&= {1 \over 2} e^{B'} \Big(
	\mathcal{I}_{J'_+}
	+ \mathcal{I}_{J'_-}
	+ \mathcal{I}_{\omega'_+}
	- \mathcal{I}_{\omega'_-} \Big) e^{-B'}, 
\label{eq:T-dual_genKaehler_plus_2} \\
\mathcal{J}'_-
&= {1 \over 2} e^{B'} \Big(
	\mathcal{I}_{J'_+}
	- \mathcal{I}_{J'_-}
	+ \mathcal{I}_{\omega'_+}
	+ \mathcal{I}_{\omega'_-} \Big) e^{-B'},
\label{eq:T-dual_genKaehler_minus_2}
\end{align}
where $J'_\pm$, $\omega'_\pm = - g' J'_{\pm}$, and $B'$ are components after the
T-duality transformation.
In particular, the metric $g'_{\mu \nu}$ and 
the $B$-field $B'_{\mu \nu}$ are given by the Buscher rule \eqref{eq:Buscher_rule}.
We stress that 
$\mathcal{I}_{J'_\pm} \neq h_y \mathcal{I}_{J_\pm} h_y$, 
$\mathcal{I}_{\omega'_\pm} \neq h_y \mathcal{I}_{\omega_\pm} h_y$, and
$e^{B'} \neq h_y e^B h_y$ in general.

By using equations~\eqref{eq:T-dual_genKaehler_plus_1}, \eqref{eq:T-dual_genKaehler_minus_1}, 
\eqref{eq:T-dual_genKaehler_plus_2}, and
\eqref{eq:T-dual_genKaehler_minus_2}, 
we obtain 
\begin{align}
\mathcal{J}'_+ + \mathcal{J}'_-
= e^{B'} \Big( \mathcal{I}_{J'_+} + \mathcal{I}_{\omega'_+} \Big) e^{-B'}
&= h_y e^B \Big( 
	\mathcal{I}_{J_+} + \mathcal{I}_{\omega_+}  \Big) e^{-B} h_y, 
\label{eq:T-dual_genKaehler_plus_4} \\
\mathcal{J}'_+ - \mathcal{J}'_-
= e^{B'} \Big( \mathcal{I}_{J'_-} - \mathcal{I}_{\omega'_-} \Big) e^{-B'}
&= h_y e^B \Big( 
	\mathcal{I}_{J_-} - \mathcal{I}_{\omega_-} \Big) e^{-B} h_y.
\label{eq:T-dual_genKaehler_minus_4}
\end{align}
Since $\mathcal{I}_{J'_\pm}$ and $\mathcal{I}_{\omega'_\pm}$ are 
 diagonal and off-diagonal block matrices respectively, 
we can read off the explicit forms of $J'_\pm$ and $\omega'_\pm$
from the most right-hand side parameterization in the above equations.
Explicitly, by multiplying $e^{-B'}$ from the left and $e^{B'}$ from the right, 
in equations \eqref{eq:T-dual_genKaehler_plus_4} and
 \eqref{eq:T-dual_genKaehler_minus_4}, we extract the transformation rules
 for $J_{\pm}$ and $\omega_{\pm}$.

We now explore the T-duality relation 
between two different bi-hermitian geometries.
The T-duality transformation along the $y$-direction, 
from one bi-hermitian structure  
$(J_\pm, \omega_\pm)$ 
to the other 
$(J'_\pm, \omega'_\pm)$, 
is derived from the equations \eqref{eq:T-dual_genKaehler_plus_4} 
and \eqref{eq:T-dual_genKaehler_minus_4} as 
\begin{align}
\mathcal{I}_{J'_+} + \mathcal{I}_{\omega'_+} 
&= e^{-B'} h_y e^B \Big( 
	\mathcal{I}_{J_+} + \mathcal{I}_{\omega_+} 
	\Big) e^{-B} h_y e^{B'}, 
\label{eq:T-dual_genKaehler_plus_5} \\
\mathcal{I}_{J'_-} - \mathcal{I}_{\omega'_-} 
&= e^{-B'} h_y e^B \Big( 
	\mathcal{I}_{J_-} - \mathcal{I}_{\omega_-} 
	\Big) e^{-B} h_y e^{B'}.
\label{eq:T-dual_genKaehler_minus_5}
\end{align}
Here $e^{B'}$ is a matrix given by
\begin{align}
e^{B'}
&= 
\begin{pmatrix}
\delta^i{}_j & 0 & 0 & 0 \\
0 & 1 & 0 & 0 \\
B'_{ij} & B'_{iy} & \delta_i{}^j & 0 \\
B'_{yj} & 0 & 0 & 1
\end{pmatrix}
= 
\begin{pmatrix}
\delta^i{}_j & 0 & 0 & 0 \\
0 & 1 & 0 & 0 \\
B_{ij} - B_{iy} g^{-1}_{yy} g_{yj} - g_{iy} g^{-1}_{yy} B_{yj} & g_{iy} g_{yy}^{-1} & \delta_i{}^j & 0 \\
- g_{yy}^{-1} g_{yj} & 0 & 0 & 1
\end{pmatrix},
\quad 
(i,j \not= y),
\label{eq:Buscher_eB}
\end{align}
where we have used the Buscher rule \eqref{eq:Buscher_rule}.
We decompose the components of the bi-complex structure 
and the fundamental two-form in the 
bi-hermitian geometry as
\begin{align}
(J_{\pm})^{\mu}{}_{\nu} 
&= 
\begin{pmatrix}
(J_\pm)^i{}_j & (J_\pm)^i{}_y \\
(J_\pm)^y{}_j & (J_\pm)^y{}_y
\end{pmatrix},
\qquad
(\omega_{\pm})_{\mu \nu}
= 
\begin{pmatrix}
(\omega_\pm)_{ij} & (\omega_\pm)_{iy} \\
(\omega_\pm)_{yj} & 0
\end{pmatrix}.
\end{align}
Then $\mathcal{I}_{J_\pm}$ and $\mathcal{I}_{\omega_\pm}$ are expressed as 
\begin{align}
\mathcal{I}_{J_\pm} 
=
\begin{pmatrix}
(J_\pm)^i{}_j & (J_\pm)^i{}_y & 0 & 0 \\
(J_\pm)^y{}_j & (J_\pm)^y{}_y & 0 & 0 \\
0 & 0 & - (J^*_\pm)_i{}^j &  - (J^*_\pm){}_i{}^y \\
0 & 0 & - (J^*_\pm)_y{}^j & - (J^*_\pm)_y{}^y
\end{pmatrix}, \quad
\mathcal{I}_{\omega_\pm}
= 
\begin{pmatrix}
0 & 0 & - (\omega^{-1}_\pm)^{ij} & - (\omega^{-1}_\pm)^{iy} \\
0 & 0 & - (\omega^{-1}_\pm)^{yj} & 0 \\
(\omega_\pm)_{ij} & (\omega_\pm)_{iy} & 0 & 0 \\
(\omega_\pm)_{yj} & 0 & 0 & 0
\end{pmatrix}.
\label{eq:generalized_complex_matrices}
\end{align}
Using the parameterizations \eqref{eq:Buscher_eB} and \eqref{eq:generalized_complex_matrices}, we find 
\begin{align}
&e^{-B'} h_y e^B \mathcal{I}_{J_\pm} e^{-B} h_y e^{B'}
\notag \\
&= \scalebox{.9}{$\renewcommand{\arraystretch}{1.2}
\begin{pmatrix}
(J_\pm)^i{}_j - (J_\pm)^i{}_4 g_{yj} g_{yy}^{-1} & 0 & 0 & (J_\pm)^i{}_y \\
B_{yk} ((J_\pm)^k{}_j - (J_\pm)^k{}_y g_{yj} g_{yy}^{-1})
	& 0 
	& - (J^*_\pm)_y{}^j 
	& B_{yk} + (J^*_\pm)_y{}^l B_{ly} \\
((\omega_\pm)_{iy} B_{yj} - B_{iy} (\omega_\pm)_{yj}) g_{yy}^{-1}
	& - (\omega_\pm)_{iy} g_{yy}^{-1} 
	& g_{iy} g_{yy}^{-1} (J_\pm^*)_y{}^j - (J_\pm^*)_i{}^j 
	& ((J_\pm^*)_i{}^l - g_{iy} (J_\pm^*)_y{}^l g_{yy}^{-1}) B_{ly} \\
- g_{yy}^{-1} (\omega_\pm)_{yj} 
	& 0 & 0 & 0
\end{pmatrix}
$},
\label{eq:calc_complex_structure_last}
\\
&e^{-B'} h_y e^B \mathcal{I}_{\omega_\pm} e^{-B} h_y e^{B'}
\notag \\
&= \scalebox{.9}{$\renewcommand{\arraystretch}{1.2}
\begin{pmatrix}
(J_\pm)^i{}_y B_{yj} g_{yy}^{-1}
	& - (J_\pm)^i{}_y g_{yy}^{-1}
	& - (\omega_\pm^{-1})^{ij} 
	& (\omega_\pm^{-1})^{il} B_{ly} \\
B_{yk} (J_\pm)^k{}_y B_{yj} g_{yy}^{-1} 
+ (\omega_\pm)_{yj}  
	& - B_{yk} (J_\pm)^k{}_y g_{yy}^{-1}
	& - B_{yk} (\omega_\pm^{-1})^{kj} 
	& B_{yk} (\omega_\pm^{-1})^{kl} B_{ly} \\
(\omega_\pm)_{ij} - (g_{iy} (\omega_\pm)_{yj} + (\omega_\pm)_{iy} g_{yj}) g_{yy}^{-1} 
	& 0
	& B_{iy} (J_\pm^*)_y{}^j g_{yy}^{-1} 
	& (\omega_\pm)_{iy} - B_{iy} (J_\pm^*)_y{}^l B_{ly} g_{yy}^{-1} \\
0
	& 0
	& (J_\pm^*)_y{}^j g_{yy}^{-1} 
	& - (J_\pm^*)_y{}^l B_{ly} g_{yy}^{-1}
\end{pmatrix}
$}, 
\label{eq:calc_Kaehler_form_last}
\end{align}
where we have used the compatibility condition 
$J_\pm^* g = \omega_\pm$.
Substituting the results of the calculations \eqref{eq:calc_complex_structure_last} 
and \eqref{eq:calc_Kaehler_form_last} into the equations \eqref{eq:T-dual_genKaehler_plus_5}
and \eqref{eq:T-dual_genKaehler_minus_5}, and comparing them with the parametrization in
\begin{align}
\mathcal{I}_{J'_\pm} \pm \mathcal{I}_{\omega'_\pm} 
&= 
\begin{pmatrix}
J'_\pm & \mp \omega^{\prime -1}_\pm \\
\pm \omega'_\pm & - J^{\prime *}_\pm
\end{pmatrix},
\end{align}
we obtain the T-duality transformation rule 
of the bi-hermitian structures from $(J_\pm, \omega_\pm)$ to $(J'_\pm, \omega'_\pm)$;
\begin{alignat}{2}
(J'_\pm)^i{}_j
&= (J_\pm)^i{}_j - {(J_\pm)^i{}_y (g_{yj} \mp B_{yj}) \over g_{yy}},
\qquad &
(J'_\pm)^i{}_y
&= \mp {(J_\pm)^i{}_y \over g_{yy}}, 
\notag \\
(J'_\pm)^y{}_j
&= \pm (\omega_\pm)_{yj} 
	+ B_{yk} \left( (J_\pm)^k{}_j - {(J_\pm)^k{}_y (g_{yj} \mp B_{yj}) \over g_{yy}} \right) , 
\qquad &
(J'_\pm)^y{}_y
&= \mp {B_{yk} (J_\pm)^k{}_y \over g_{yy}},
\notag \\
(\omega'_\pm)_{ij}
&= (\omega_\pm)_{ij} 
	- {(\omega_\pm)_{iy} (g_{yj} \mp B_{yj}) + (g_{iy} \pm B_{iy}) (\omega_\pm)_{yj} \over g_{yy}}, 
\qquad &
(\omega'_\pm)_{iy}
&= \mp {(\omega_\pm)_{iy} \over g_{yy}}.
\label{eq:Buscher_rules_for_J_and_omega_bi-hermitian}
\end{alignat}
This exhibits the Buscher-like T-duality rule between two different bi-hermitian structures.
The formula, derived from the doubled formalism, precisely agrees with the
ones discussed in the context of worldsheet supersymmetry
\cite{Ivanov:1994ec, Hassan:1994mq, Bakas:1995hc, Hassan:1995je}.
We find that an analogous formula holds for bi-hypercomplex cases.
We note that an alternative Buscher-like rule for the components in the
generalized almost complex structures is discussed in \cite{Persson:2006rd}.

We focus on the relation between a K\"{a}hler geometry and a bi-hermitian geometry. 
In an (almost) K\"{a}hler geometry, we have $B = 0$ 
and the two (almost) complex structures $J_+$ and $J_-$ 
coincide with each other;
$J_+ = J_- = J$, $\omega_+ = \omega_- = \omega$. 
Then, the Buscher-like T-duality rule from the K\"{a}hler structure $(J,\omega)$ 
to the bi-hermitian structure $(J'_\pm, \omega'_\pm)$ is given by
\begin{alignat}{4}
(J'_\pm)^i{}_j
&= J^i{}_j - {J^i{}_y g_{yj} \over g_{yy}},
\qquad &
(J'_\pm)^i{}_y
&= \mp {J^i{}_y \over g_{yy}},
\qquad &
(J'_\pm)^y{}_j
&= \pm \omega_{yj},
\qquad &
(J'_\pm)^y{}_y
&= 0,
\notag \\
(\omega'_\pm)_{ij}
&= \omega_{ij} - {\omega_{iy} g_{yj} + g_{iy} \omega_{yj} \over g_{yy}}, 
\qquad &
(\omega'_\pm)_{iy}
&= \mp {\omega_{iy} \over g_{yy}}.
&&&&
\label{eq:Buscher_rules_for_J_and_omega}
\end{alignat}
The T-duality rule from the hyperk\"{a}hler $(J_a, \omega_a),\, (a =1,2,3)$
to the bi-hypercomplex $(J_{a,\pm},\omega_{a,\pm})$ is analogous.

We now calculate the square of the almost complex structures
after the T-duality transformation. 
From the equation~\eqref{eq:Buscher_rules_for_J_and_omega}, 
the square of $J'_\pm$ can be calculated as 
\begin{align}
(J'_\pm)^2
&= 
\begin{pmatrix}
J^i{}_k - {J^i{}_y g_{yk} \over g_{yy}} & \mp {J^i{}_y \over g_{yy}} \\
\pm \omega_{yk} & 0
\end{pmatrix}
\begin{pmatrix}
J^k{}_j - {J^k{}_y g_{yj} \over g_{yy}} & \mp {J^k{}_y \over g_{yy}} \\
\pm \omega_{yj} & 0
\end{pmatrix}
\notag \\
&= 
\scalebox{.9}{$
{\renewcommand{\arraystretch}{1.4}
\begin{pmatrix}
J^i{}_k J^k{}_j - {1 \over g_{yy}} \big( J^i{}_k J^k{}_y g_{yj} + J^i{}_y g_{yk} J^k{}_j \big)
	+ {1 \over g_{yy}^2} J^i{}_y g_{yk} J^k{}_y g_{yj} 
	- {J^i{}_y \over g_{yy}} \omega_{yj}
	& \mp J^i{}_k {J^k{}_y \over g_{yy}} \pm {J^i{}_y g_{yk} J^k{}_y \over g_{yy}^2} \\
\pm \omega_{yk} J^k{}_j \mp \omega_{yk} {J^k{}_y g_{yj} \over g_{yy}}
	& - {\omega_{yk} J^k{}_y \over g_{yy}}
\end{pmatrix}}
$}.
\end{align}
By the compatibility conditions $\omega = - gJ$ and $g = \omega J$, 
we have
\begin{align}
(J'_\pm)^2
&= 
{\renewcommand{\arraystretch}{1.4}
\begin{pmatrix}
J^i{}_k J^k{}_j - {1 \over g_{yy}} J^i{}_k J^k{}_y g_{yj} 
	- {1 \over g_{yy}} J^i{}_y J^y{}_y g_{yj} 
	+ J^i{}_y J^y{}_j
	& \mp J^i{}_k {J^k{}_y \over g_{yy}} \mp {J^i{}_y J^y{}_y \over g_{yy}} \\
0 & - 1
\end{pmatrix}}.
\end{align}
Furthermore, by using the condition $J^2 = -1$,
we find 
\begin{align}
(J'_\pm)^2
&= 
\begin{pmatrix}
- \delta^i{}_j & 0 \\
0 & - 1
\end{pmatrix}
= -1. 
\end{align}
This shows that both of $J'_\pm$, after the T-duality transformation,
satisfy the property of the almost complex structure.

We finally examine the commutativity of $J'_+$ and $J'_-$. 
The product of $J'_+$ and $J'_-$ is
\begin{align}
J'_\pm J'_\mp 
&= 
\begin{pmatrix}
J^i{}_k - {J^i{}_y g_{yk} \over g_{yy}} & \mp {J^i{}_y \over g_{yy}} \\
\pm \omega_{yk} & 0
\end{pmatrix}
\begin{pmatrix}
J^k{}_j - {J^k{}_y g_{yj} \over g_{yy}} & \pm {J^k{}_y \over g_{yy}} \\
\mp \omega_{yj} & 0
\end{pmatrix}
\notag \\
&= 
\scalebox{.9}{$
{\renewcommand{\arraystretch}{1.4}
\begin{pmatrix}
J^i{}_k J^k{}_j - {1 \over g_{yy}} \big( J^i{}_k J^k{}_y g_{yj} + J^i{}_y g_{yk} J^k{}_j \big)
	+ {1 \over g_{yy}^2} J^i{}_y g_{yk} J^k{}_y g_{yj} 
	+ {J^i{}_y \over g_{yy}} \omega_{yj}
	& \pm J^i{}_k {J^k{}_y \over g_{yy}} \mp {J^i{}_y g_{yk} J^k{}_y \over g_{yy}^2} \\
\pm \omega_{yk} J^k{}_j \mp \omega_{yk} {J^k{}_y g_{yj} \over g_{yy}}
	& {\omega_{yk} J^k{}_y \over g_{yy}}
\end{pmatrix}}
$}.
\end{align}
By using the compatibility condition $g = \omega J$, $\omega = - gJ$, and $J^2 = -1$, 
we obtain 
\begin{align}
J'_+ J'_- 
&= J'_- J'_+ =
\begin{pmatrix}
J^i{}_k J^k{}_j 
	+ {1 \over g_{yy}} J^i{}_y \big( \omega_{yj} - g_{yk} J^k{}_j \big)
	& 0 \\
0 & 1
\end{pmatrix}.
\end{align}
Then the relation $[J'_+, J'_-] = 0$ holds.
Therefore we find that the commutativity of the bi-hermitian structures 
is preserved by T-duality transformations.

\section{T-duality from KK- to H-monopole geometries} \label{section:HM_KKM}
It is useful to introduce a concrete example in order to confirm the
formulas in the previous section.
In this section, we consider T-duality between the KK5- and the
NS5-branes in type II string theories.
They are BPS solutions to type II supergravities preserving sixteen supersymmetries.
A fundamental string propagating on these backgrounds preserves eight supersymmetries and 
its worldsheet theory has $\mathcal{N} = (4,4)$ supersymmetry.
Then the geometries admit at least one hyperk\"{a}hler structure.
We note that the KK5- and the NS5-branes are incorporated into a single solution in the doubled space $\mathcal{M}$
\cite{Berman:2014jsa}.
This is known as the DFT monopole and the KK5- and the NS5-branes appear in a particular $O(D,D)$ frame.
Then the DFT monopole geometry should have a generalized hypercomplex structure that encompasses 
bi-hypercomplex structures of spacetimes.
The procedure discussed in the previous section is directly utilized to relate 
the bi-hypercomplex (hyperk\"{a}hler) structures of the KK5- and the NS5-brane geometries.
We elucidate this in the following.

The four-dimensional transverse geometry of the KK5-brane 
(also known as the KK-monopole) is given by
\begin{align}
{\dop}s^2
&= H \, {\dop}x_{123}^2 + H^{-1} \big( {\dop}x^4 + A_i \, {\dop}x^i
 \big)^2, \qquad (i = 1,2,3),
\label{eq:KK-monopole}
\end{align}
where $H$ is a harmonic function in the flat three dimensions $(x^1,x^2,x^3)$
and $A_i$ ($i=1,2,3$) is a vector potential satisfying the monopole equation 
${\dop} A = \hat{*}_3 {\dop}H$. 
Here $\hat{*}_3$ is the Hodge star operator defined in
the $(x^1,x^2,x^3)$ space. The isometry lies in the $x^4$-direction.
This is the Euclidean Taub-NUT space admitting 
three complex structures $J_a$ ($a=1,2,3$).

For later convenience we write the metric components;  
\begin{align}
g_{\mu\nu} 
&= H^{-1}
\begin{pmatrix}
H^2 + A_1^2 & A_1 A_2 & A_1 A_3 & A_1 \\
A_1 A_2 & H^2 + A_2^2 & A_2 A_3 & A_2 \\
A_1 A_3 & A_2 A_3 & H^2 + A_3^2 & A_3 \\
A_1 & A_2 & A_3 & 1
\end{pmatrix},
\label{eq:Taub-NUT_metric_matrix}
\end{align}
and the three fundamental two-forms
on the Taub-NUT space;
\begin{align}
\omega_1
&= {\dop}x^1 \wedge \big( {\dop}x^4 + A \big) + H \, {\dop}x^2 \wedge {\dop}x^3, \notag \\
\omega_2 
&= {\dop}x^2 \wedge \big( {\dop}x^4 + A \big) + H \, {\dop}x^3 \wedge {\dop}x^1, \notag \\
\omega_3 
&= {\dop}x^3 \wedge \big( {\dop}x^4 + A \big) + H \, {\dop}x^1 \wedge {\dop}x^2. 
\label{eq:Kahler_form_Taub-NUT}
\end{align}
They satisfy the definition of the hyperk\"{a}hler structure ${\dop}\omega_i = 0$ 
when the monopole equation 
 ${\dop}A = \hat{*}_3 {\dop}H$ is imposed.
Then the explicit form of three complex structures
is given by
\begin{align}
(J_1)^\mu{}_\nu
&= - H^{-1}
\begin{pmatrix}
A_1 & A_2 & A_3 & 1 \\
0 & 0 & H & 0 \\
0 & - H & 0 & 0 \\
- H^2 - A_1^2 
	& - A_1 A_2 + A_3 H 
	& - A_1 A_3 - A_2 H 
	& - A_1
\end{pmatrix},
\notag \\%[.3zh]
(J_2)^\mu{}_\nu
&= - H^{-1}
\begin{pmatrix}
0 & 0 & - H & 0 \\
A_1 & A_2 & A_3 & 1 \\
H & 0 & 0 & 0 \\
- A_1 A_2 - A_3 H 
	& - H^2 - A_2^2 
	& H A_1 - A_2 A_3 
	& - A_2
\end{pmatrix},
\notag \\%[.3zh]
(J_3)^\mu{}_\nu
&= - H^{-1}
\begin{pmatrix}
0 & H & 0 & 0 \\
- H & 0 & 0 & 0 \\
A_1 & A_2 & A_3 & 1 \\
A_2 H - A_1 A_3 
	& - H A_1 - A_2 A_3 
	& - H^2 - A_3^2 
	& - A_3
\end{pmatrix}. 
\label{eq:complex_structure_on_Taub-NUT_matrix}
\end{align}
One finds that this expression satisfies the quaternion algebra 
$J_a J_b = - \delta_{ab} + \epsilon_{abc} J_c$.

Now we perform the T-duality transformation along the $x^4$-direction.
By applying the Buscher rule~\eqref{eq:Buscher_rule} 
to the metric \eqref{eq:KK-monopole}, we obtain
\begin{align}
g'_{\mu\nu}
= 
{\renewcommand{\arraystretch}{.85}
\begin{pmatrix}
H & 0 & 0 & 0 \\
0 & H & 0 & 0 \\
0 & 0 & H & 0 \\
0 & 0 & 0 & H 
\end{pmatrix}}, 
\qquad 
B'_{\mu\nu} 
= 
{\renewcommand{\arraystretch}{.85}
\begin{pmatrix}
0 & 0 & 0 & A_1 \\
0 & 0 & 0 & A_2 \\
0 & 0 & 0 & A_3 \\
- A_1 & - A_2 & - A_3 & 0
\end{pmatrix}},
\qquad 
e^{2\phi'} = H.
\end{align}
The geometry described by this metric, the $B$-field and the dilaton 
gives the four-dimensional transverse space of the smeared NS5-brane, which is also called the H-monopole.

We now examine the bi-hypercomplex structures of the H-monopole.
By applying the T-duality rule~\eqref{eq:Buscher_rules_for_J_and_omega} to 
the fundamental forms~\eqref{eq:Kahler_form_Taub-NUT} 
and the three complex structures~\eqref{eq:complex_structure_on_Taub-NUT_matrix},
we evaluate the bi-hypercomplex structures 
$(J'_{a,\pm}, \omega'_{a,\pm})$ on the H-monopole geometry. The result is 
\begin{alignat}{3}
J'_{1,+}
&= 
{\renewcommand{\arraystretch}{.85}
\begin{pmatrix}
0 & 0 & 0 & 1 \\
0 & 0 & -1 & 0 \\
0 & 1 & 0 & 0 \\
-1 & 0 & 0 & 0 
\end{pmatrix}}, &\quad
J'_{2,+}
&= 
{\renewcommand{\arraystretch}{.85}
\begin{pmatrix}
0 & 0 & 1 & 0 \\
0 & 0 & 0 & 1 \\
-1 & 0 & 0 & 0 \\
0 & -1 & 0 & 0 
\end{pmatrix}}, &\quad
J'_{3,+}
&= 
{\renewcommand{\arraystretch}{.85}
\begin{pmatrix}
0 & -1 & 0 & 0 \\
1 & 0 & 0 & 0 \\
0 & 0 & 0 & 1 \\
0 & 0 & -1 & 0 
\end{pmatrix}}, 
\notag \\
J'_{1,-}
&= 
{\renewcommand{\arraystretch}{.85}
\begin{pmatrix}
0 & 0 & 0 & -1 \\
0 & 0 & -1 & 0 \\
0 & 1 & 0 & 0 \\
1 & 0 & 0 & 0 
\end{pmatrix}}, &\quad
J'_{2,-}
&= 
{\renewcommand{\arraystretch}{.85}
\begin{pmatrix}
0 & 0 & 1 & 0 \\
0 & 0 & 0 & -1 \\
-1 & 0 & 0 & 0 \\
0 & 1 & 0 & 0 
\end{pmatrix}}, &\quad
J'_{3,-}
&= 
{\renewcommand{\arraystretch}{.85}
\begin{pmatrix}
0 & -1 & 0 & 0 \\
1 & 0 & 0 & 0 \\
0 & 0 & 0 & -1 \\
0 & 0 & 1 & 0 
\end{pmatrix}}, 
\label{eq:H-monopole_bi-hyper-complex_strc}
\\%[.5zh]
\omega'_{1,+}
&= 
{\renewcommand{\arraystretch}{.85}
\begin{pmatrix}
0 & 0 & 0 & -H \\
0 & 0 & H & 0 \\
0 & -H & 0 & 0 \\
H & 0 & 0 & 0 
\end{pmatrix}}, &\quad
\omega'_{2,+}
&= 
{\renewcommand{\arraystretch}{.85}
\begin{pmatrix}
0 & 0 & -H & 0 \\
0 & 0 & 0 & -H \\
H & 0 & 0 & 0 \\
0 & H & 0 & 0 
\end{pmatrix}}, &\quad
\omega'_{3,+}
&= 
{\renewcommand{\arraystretch}{.85}
\begin{pmatrix}
0 & H & 0 & 0 \\
-H & 0 & 0 & 0 \\
0 & 0 & 0 & -H \\
0 & 0 & H & 0 
\end{pmatrix}}, 
\notag \\
\omega'_{1,-}
&= 
{\renewcommand{\arraystretch}{.85}
\begin{pmatrix}
0 & 0 & 0 & H \\
0 & 0 & H & 0 \\
0 & -H & 0 & 0 \\
-H & 0 & 0 & 0 
\end{pmatrix}}, 
&\quad
\omega'_{2,-}
&=
{\renewcommand{\arraystretch}{.85}
\begin{pmatrix}
0 & 0 & -H & 0 \\
0 & 0 & 0 & H \\
H & 0 & 0 & 0 \\
0 & -H & 0 & 0 
\end{pmatrix}}, 
&\quad
\omega'_{3,-}
&=
{\renewcommand{\arraystretch}{.85}
\begin{pmatrix}
0 & H & 0 & 0 \\
-H & 0 & 0 & 0 \\
0 & 0 & 0 & H \\
0 & 0 & -H & 0 
\end{pmatrix}}.
\label{eq:dual_two-form}
\end{alignat}
We find the complete agreement among the expressions
\eqref{eq:H-monopole_bi-hyper-complex_strc}, \eqref{eq:dual_two-form}
and the ones in the literature \cite{Papadopoulos:2000iv}.
We also confirm that these structures satisfy the compatibility condition
\begin{align}
\omega'_{a,\pm} J'_{a,\pm} 
&= g',
\quad (\text{$a$: no sum}).
\end{align}
Furthermore, we find that $J'_{a,\pm}$ are covariantly constant in the
following sense; 
\begin{align}
\nabla^{(+)}_\rho (J'_{a,+})^\mu{}_\nu
&= \nabla_\rho (J'_{a,+})^\mu{}_\nu
+ {1 \over 2} \Big( 
	(J'_{a,+})^\mu{}_\sigma g^{\sigma\alpha} H'_{\rho\nu\alpha} 
	- (J'_{a,+})^\sigma{}_\nu g^{\mu\alpha} H'_{\rho\sigma\alpha} \Big)
= 0,
\notag \\
\nabla^{(-)}_\rho (J'_{a,-})^\mu{}_\nu
&= \nabla_\rho (J'_{a,-})^\mu{}_\nu 
- {1 \over 2} \Big( 
	(J'_{a,-})^\mu{}_\sigma g^{\sigma\alpha} H'_{\rho\nu\alpha} 
	- (J'_{a,-})^\sigma{}_\nu g^{\mu\alpha} H'_{\rho\sigma\alpha} \Big)
= 0,
\end{align}
where $\nabla$ involves
the Levi-Civita connection compatible with the metric 
$g'_{\mu \nu}$, and $H'_{\mu\nu\rho}$ is the field strength of 
$B'_{\mu \nu}$. 
They are nothing but the conditions \eqref{eq:covariantly_constant}
derived from the $\mathcal{N} = (4,4)$ sigma models in Section \ref{section:NLSM}.

\section{Worldsheet instantons and T-duality} \label{section:instantons}
We have established the explicit T-duality relations between 
the K\"{a}hler (hyperk\"{a}hler) and the bi-hermitian (bi-hypercomplex) geometries.
In particular, the T-duality relation between the 
KK- and the H-monopole geometries are now apparent.
An important notion that depends on these geometric structures is the
worldsheet instantons in string theory \cite{Wen:1985jz}.
In this section, we discuss its T-duality relation.

The worldsheet instanton equation is derived by the Bogomol'nyi completion of the Euclidean worldsheet action of string;
\begin{align}
S_{\mathrm{E}} =& \ 
\frac{T}{2} 
\int_{\Sigma} \! {\dop}^2 \xi \, 
\sqrt{h} g_{\mu \nu} h^{ab} \del_a X^{\mu} \del_b X^{\nu}
\notag \\
=& \ 
\frac{T}{4} 
\int_{\Sigma} \! {\dop}^2 \xi \, \sqrt{h} 
h^{ab} g_{\mu \nu} 
\left(
\del_a X^{\mu} \pm J^{\mu} {}_{\rho} \epsilon_{ac} \del^c X^{\rho} 
\right)
\left(
\del_b X^{\nu} \pm J^{\nu} {}_{\sigma} \epsilon_{bd} \del^d X^{\sigma}
\right).
\notag \\
& 
\mp 
\frac{T}{2}
\int_{\Sigma} \! {\dop}^2 \xi \, \sqrt{h} 
g_{\mu \nu} h^{ab} \epsilon_{ac} J^{\mu} {}_{\rho} \del^c X^{\rho}
 \del_b X^{\nu},
\end{align}
where $T$ is the string tension, $h_{ab} \, (a,b=1,2)$ is the metric of the worldsheet $\Sigma$ and $\epsilon_{ab}$ is the antisymmetric tensor on $\Sigma$.
$g_{\mu \nu}$ and $J^{\mu} {}_{\nu}$ are the metric and 
the complex structure of a K\"{a}hler geometry $M$. 
The action is bounded from below when the following equation is satisfied;
\begin{align}
\del_a X^{\mu} \pm J^{\mu} {}_{\rho} \epsilon_{ac} \del^c X^{\rho} = 0.
\end{align}
These are the worldsheet instanton equations.
(Anti-)holomorphic maps $X: \Sigma \to C^2$ are solutions to these
equations and they are called the worldsheet instantons.
Here $C^2$ is a two-cycle in the spacetime $M$.
The action is given by
\begin{align}
S_{\mathrm{E}} = \pm \int_{C^2} \! \omega_J. 
\end{align}
Here we employ a convention $T=2$ and 
$(\omega_J)_{\mu \nu} = - g_{\mu \rho} J^{\rho}{}_{\nu}$ 
is the K\"{a}hler form on $M$.

Now we consider the T-duality transformation of all the materials by 
introducing an isometry for the background along the $X^y$-direction,
and making it be gauged.
We introduce the Lagrange multiplier $\tilde{X}^y$ in the worldsheet
action to ensure the vanishing gauge
field strength and then integrate out $X^y$.
The procedure results in the T-dualized background
\eqref{eq:Buscher_rule}, which 
entails 
the bi-hermitian 
structures.
This together with the transformation rules
\eqref{eq:Buscher_rules_for_J_and_omega} for geometric structures leads 
to the instanton equations in 
the bi-hermitian geometry;
\begin{align}
\del_a X^{\mu} \pm J^{\prime \mu}_{\pm} {}_{\nu} \epsilon_{ac} \del^c X^{\rho}
 = 0,
\end{align}
where $X^{\mu} = (X^i, \tilde{X}^y), \, (i \not= y)$.
It is noteworthy that there is a one-to-two correspondence between the
instantons in the K\"{a}hler and the bi-hermitian geometries.
Among other things, we have two options for the topological terms associated with instantons;
\begin{align}
S'_{\mathrm{E},+} = \pm \int_{C^2} \omega_{J'_+} + i \int_{C^2} B',
\qquad
S'_{\mathrm{E},-} = \pm \int_{C^2} \omega_{J'_-} + i \int_{C^2} B'.
\end{align}

The T-duality between the KK- and the H-monopole geometries goes in this class.
Indeed, the worldsheet instanton effects of these geometries are
intensively studied \cite{Gregory:1997te, Tong:2002rq, Harvey:2005ab, 
Okuyama:2005gx, Kimura:2018hph}.
A well-known fact is that two-cycles that support instantons exist 
in the KK-monopole geometry but there are no such cycles in the
T-dualized H-monopole side\footnote{
Strictly speaking, two-cycles exist only in the multi-centred Taub-NUT spaces.
For the single-centred Taub-NUT space, we can define disk instantons
instead, for which strings wrap 
on a cigar-type geometry. See \cite{Okuyama:2005gx, Kimura:2018hph} for details.
}.
It has been discussed that instantons in the
H-monopole geometry are 
point-like 
instantons with vanishing two-cycles \cite{Tong:2002rq, Harvey:2005ab}.
However, no explicit proof of this proposal has been known.
The explicit T-duality relations of complex structures and 
this splitting effect of instantons would play an important role to
understand this puzzle.

\section{Conclusion and discussions} \label{section:conclusion}

In this paper, we studied detailed relations among 
the K\"{a}hler (hyperk\"{a}hler) structures, the bi-hermitian (bi-hypercomplex) structures, 
generalized or doubled geometries and T-duality.
Although these notions were individually studied in various contexts, their
comprehensive connections have been overlooked in the literature.

The bi-hermitian and the bi-hypercomplex geometries are the target
spaces of two-dimensional $\mathcal{N} = (2,2)$ and $\mathcal{N} = (4,4)$ 
non-linear sigma models with twisted chiral (or hyper) multiplets.
Since the chiral and the twisted chiral multiplets are interchanged by
T-duality transformation in the two-dimensional theories, 
there are explicit T-duality relations 
between these geometries.
Indeed, this has been intensively studied in the context of supersymmetric non-linear sigma models.

On the other hand, the 
complex and the bi-hermitian structures are expressed in terms of generalized or 
doubled geometries where T-duality symmetry is more apparent.
The bi-hermitian and the bi-hypercomplex structures are embedded into
the generalized complex and the generalized hyperk\"{a}hler structures,
respectively.
The T-duality transformation is implemented by the $O(D,D)$ rotations in
the doubled formalism.
We exhibited how to extract the transformation rule for the bi-hermitian (bi-hypercomplex) structures
in the doubled formalism and wrote down the explicit Buscher-like formula in the component form.
The result precisely reproduced the relations derived in the supersymmetric sigma models \cite{Ivanov:1994ec, Hassan:1994mq, Bakas:1995hc, Hassan:1995je}.
Although the formula itself has been known, the cumbersome procedures
deriving the formula in sigma models or mathematically rigorous
treatments \cite{Cavalcanti:2011wu} are drastically simplified in the doubled formalism.
We also stress that our technique is available not only for geometries realized by supersymmetric sigma models 
but also for any geometries admitting appropriate bi-hermitian (bi-hypercomplex) structures.
The bi-hermitian (bi-hypercomplex) structures of spacetimes are embedded in geometrical structures of the doubled space $\mathcal{M}$ and their 
transformations are just changes of basis for endomorphisms on $T\mathcal{M}$.
Therefore our procedure is easily generalized to general $O(D,D)$ transformations rather than that based on the factorized T-duality developed 
in the context of sigma models.

The derived formula helps us to find how the geometric structures of spacetimes are interchanged via T-duality.
Among other things, we showed the relations between the hyperk\"{a}hler
structure on the KK-monopole and the bi-hypercomplex structure on the H-monopole.
By applying our formula to the KK-monopole geometry, we re-derived the
bi-hypercomplex structure of the H-monopole first found in \cite{Papadopoulos:2000iv}.
As a byproduct, we discussed the T-duality between the worldsheet
instantons in the KK- and the H-monopole geometries.
We found that there is a one-to-two correspondence between them.
Since the worldsheet instantons cause drastic changes of spacetime
geometries, we expect that our result would help to understand 
non-trivial relationships between T-dualized geometries.

We exhibited the bi-hypercomplex nature of the H-monopole (smeared NS5-brane)
geometry with non-trivial torsion $T = H = {\dop}B$.
By performing a T-duality transformation
along a certain direction, 
we obtain the KK-monopole described by the hyperk\"{a}hler Taub-NUT geometry 
without torsion.
When we introduce an additional isometry and perform another T-duality
transformation along this direction, we obtain the geometry of the exotic $5^2_2$-brane
\cite{Obers:1998fb}. This geometry is again torsionful and preserve the
same supersymmetry as in the H- and the KK-monopoles \cite{Kimura:2014bea}.
Therefore it is expected that the $5^2_2$-brane geometry admits
bi-hypercomplex structures, at least locally\footnote{
Notice that the single $5^2_2$-brane is not well-defined as a stand-alone object \cite{deBoer:2010ud}. In order to discuss the globally well-defined structure of the system, we should incorporate two additional branes \cite{Kikuchi:2012za}.}.
This is also anticipated from the fact that the $5^2_2$-brane appears in a particular $O(D,D)$ frame of the DFT monopole \cite{Bakhmatov:2016kfn}.
We again stress that our Buscher-like rule \eqref{eq:Buscher_rules_for_J_and_omega} derived from the doubled formalism 
is applicable, not limited to geometries realized by supersymmetric sigma models, to any geometries solving DFT (and hence supergravities).
Apart from this fact, a
sigma model description of the $5^2_2$-brane geometry \cite{Kimura:2013fda},
and the instantons in the $5^2_2$-brane geometry \cite{Kimura:2013zva}
are studied. It would be interesting to study their relations to
the bi-hermitian geometry.

In order to get a better understanding of T-duality nature of geometric structures,
it is useful to work in the doubled formalism.
Indeed, the $O(D,D)$ and spacetime structures are incorporated into the Born
geometry in a T-duality covariant way \cite{Freidel:2017yuv, Freidel:2018tkj, Freidel:2013zga}.
The T-dual properties of worldsheet instantons are best examined in the
Born sigma model \cite{Marotta:2019eqc}.
We found that the generalized K\"{a}hler and the generalised hyperk\"{a}hler
structures satisfy the algebras of the hypercomplex numbers.
This implies an interesting mathematical property of the complex
structures in the doubled space.
We will report these relations in the near future.

%%%%%%%%%%%%%%%%%%%%%%%%%%%%%%%%%%%
\subsection*{Acknowledgments}
We would like to thank S.F.~Hassan and K.~Sfetsos for notifying their works and related literature on the
T-duality transformation of complex structures.
The work of T.K. and S.S. is supported in part by Grant-in-Aid for Scientific Research (C),
JSPS KAKENHI Grant Number JP20K03952. The work of K.S. is supported by Grant-in-Aid
for JSPS Research Fellow, JSPS KAKENHI Grant Number JP20J13957.

%%%%%%%%%%%%%%%%%%%%%%%%%%%%%%%%%%%

%%%%%%%%%%%%%%%%%%%%%%%%%%%%%%%%%%%

}
\end{document}